# The Giant Surface Charge Density of Graphene Resolved From Scanning Tunneling Microscopy and First Principles Theory

P. Xu,[1] Y. Yang,[1,2] S.D. Barber,[1] M.L. Ackerman,[1] J.K. Schoelz,[1]

I.A. Kornev,[3] S. Barraza-Lopez,[1] L. Bellaiche,[1] and P.M. Thibado[1*]

[1]*Department of Physics, The University of Arkansas, Fayetteville, Arkansas 72701, USA*

[2]*Physics Department, Nanjing University of Aeronautics and Astronautics, Nanjing 210016, China*

[3]*Laboratoire Structures, Proprietes et Modelisation des Solides, Ecole Centrale Paris, CNRS-UMR8580, Grande Voie des Vignes, 92295 Chatenay-Malabry Cedex, France*

In this work, systematic constant-bias, variable-current scanning tunneling microscopy (STM) measurements and STM simulations from density-functional theory are made, yielding critical insights into the spatial structure of electrons in graphene. A foundational comparison is drawn between graphene and graphite, showing the surface charge density of graphene to be 300% that of graphite. Furthermore, simulated STM images reveal that high-current STM better resolves graphene's honeycomb bonding structure because of a retraction which occurs in the topmost dangling bond orbitals.

PACS numbers: 68.37.Ef, 73.22.Pr, 81.05.uf, 31.15.A-

Graphene is a two-dimensional system consisting of a single plane of carbon atoms arranged in a honeycomb lattice [1,2]. In effect, it is an isolated atomic monolayer of the common semimetal graphite. However, as research efforts have shown since its isolation in 2005, graphene is hardly common. Graphene's structure causes its charge carriers to mimic massless Dirac fermions [1], thereby displaying remarkable electronic characteristics, such as extremely high mobility [3], unusually far ballistic transport [4,5], and even the half-integer quantum Hall effect [2].

Another promising feature of graphene is its enormous current-carrying capacity, on the order of $10^8$ A/cm$^2$ for nanoribbons [6]. Self-heating [7] has been found to be the limiting factor that triggers the breakdown current density. Thus, its current-carrying capacity is fundamentally linked to an extraordinary thermal conductivity [8] of about 5,000 W m$^{-1}$ K$^{-1}$. These two attributes provide evidence that graphene possesses a highly unusual electronic density of states (DOS). In fact, the DOS of graphene has been the subject of extensive theoretical and experimental studies. Angle-resolved photoemission spectroscopy [9-12] and transmission electron microscopy studies [13,14], for example, have revealed a great deal about graphene's band structure, linear energy dispersion, and electron-phonon coupling modes.

Scanning tunneling microscopy (STM) has also proven to be a powerful tool for studying the electronic and geometric structure of graphene. For example, the local density of states (LDOS) of graphene nanostructures was observed [15,16] using STM. In another study, the local carrier density [17] was examined using STM tunneling spectra. This research clarified the nature of substrate-induced spatial charge inhomogeneities, which are often limiting factors in graphene's carrier mobility [18]. Other STM studies have imaged graphene at the atomic scale, as well. All used an unusually high tunneling current, most likely made possible by graphene's unique electronic DOS. Most of these studies used 1 nA [19], some used 5 nA [20], and a few even reported using currents as high as 30 nA [21]. Studies haven't yet determined, however, the reason why such high currents can be used to image graphene. In addition, no research efforts to date have systematically reported the effect of varying the tunneling current.

In this Rapid Communication, the real-space magnitude and spatial extent of the surface charge density of graphene is determined using STM and first-principles electronic structure calculations, where variations in the tunneling current is the primary focus. Filled-state, variable-current images from both theory and experiment are compared to reveal an exceptionally large charge density which extends more than 0.2 nm away from the nuclear plane. This large surface charge density is determined to be the reason why a relatively large tunneling current is optimal for imaging the hexagonal atomic structure of graphene. Finally, graphene and graphite are examined alongside each other to directly quantify the striking differences between the two similar surfaces.

The experimental STM images were obtained using an Omicron ultrahigh-vacuum (base pressure is $10^{-10}$ Torr), low-temperature STM operated at room temperature. The graphene samples used in this study came from 2 in × 2 in sheets of 20-μm-thick copper foil [22] on which single-layer graphene had been grown via chemical vapor deposition [23,24]. A 1 cm × 1 cm square was cut from this sheet and mounted with silver paint onto a flat tantalum STM sample plate. The graphite samples used in this study were taken from a 12 mm × 12 mm by 2 mm thick piece of highly-oriented pyrolytic graphite (HOPG) [25], which was sliced into two pieces. One of the halves was mounted, and then its top layers were removed with tape to expose a fresh surface. The STM tips used to image the samples were manufactured in-house by electrochemically etching polycrystalline tungsten wire via a differential-cutoff lamella method. After etching, the tips were gently rinsed with distilled water, briefly dipped in a concentrated hydrofluoric acid solution to remove surface oxides [26], then transferred through a load-lock into the STM chamber. Numerous small-scale, filled-state STM images were acquired for both graphene and graphite using a tip bias of +0.100 V. The tunneling current was maintained at a constant value of 0.200 nA for graphite, but was varied when imaging graphene.

For the theoretical component of the study, simulated STM images of graphene and graphite were extracted from density functional theory (DFT) calculations without modeling the STM tip [27,28]. These calculations were performed within the local-density approximation to DFT using projector-augmented wave potentials [29] as implemented in the plane-wave basis set VASP [30] code. The graphene surface was modeled as a single layer using a 1×1 cell, while the graphite was modeled as a six-layer Bernal stack, also using a 1×1 cell. A cutoff energy of 400 eV and a huge 225×225×1 Monkhorst-Park k-point mesh was used to ensure proper sampling around the Dirac point. The atoms were allowed to move until all forces were less than 0.1 eV/nm, resulting in a carbon-carbon bond length of 0.142 nm and an interplanar separation of 0.334 nm. After structural relaxation, and to best replicate the experimental STM conditions, the LDOS was integrated from the Dirac point to 0.01 eV below that point. A series of constant-current simulated STM images was produced by extracting isocontour surfaces using various values for the LDOS to model different tunneling currents.

Two filled-state STM images (6 nm × 6 nm) of graphene on copper foil are shown in Fig. 1, alongside corresponding line profiles. Graphene's hexagonal structure is resolved throughout the image in Fig. 1(a), which was taken using a constant tunneling current of 1.000 nA. The inset image was cut from the central region, magnified 2.5 times, and displayed with a compressed color scale. It shows atomic resolution around a single benzene ring structure. This result demonstrates that atomic-resolution imaging in graphene can be obtained with a comparatively high tunneling current, as previously described. Note that the relatively large height variations extending across the image, especially from bottom left to top right, occur because the graphene conforms to the harsh morphology of the underlying polycrystalline copper foil. A line profile was extracted across the middle of a horizontal row – the fast scan direction –of Fig. 1(a). The line location is marked with an arrow just to the right of the image and is shown in Fig. 1(b). This profile exhibits regular sinusoidal oscillations with a corrugation amplitude of approx. 0.05 nm and a spatial period of approx. 0.3 nm.

For the second STM image, shown in Fig. 1(c), the fast scan direction is oriented vertically. The tunneling current was varied during this data run. It was set at 1.000 nA for approximately the first third of the scan, then decreased in steps to a minimum of 0.010 nA near the midpoint. The current was left at this value for a short time, then increased in steps back to 1.000 nA, where it was held for the final third of the scan. The box directly below the image divides it into regions where the tunneling current was high (H = 1.000 nA), medium (0.010 nA < M < 1.000 nA), and low (L = 0.010 nA).

Two important features should be noted in Fig. 1(c). First, the honeycomb lattice becomes noticeably less visible as the current decreases, almost disappearing at the 0.010 nA level. To show this change more clearly, three line profiles were taken along the fast scan (vertical) direction and plotted in Fig. 1(d), with offsets to prevent overlap. The topmost curve corresponds to a current of 1 nA, the middle to 0.1 nA, and the bottom to 0.01 nA. The corrugation amplitude decreases by about a factor of 10 over this range, making the oscillations within the image more difficult to resolve at low current. This is the first time variable-current features such as these have been reported; typically, bias-dependent features in STM images are seen [28].

The second notable aspect of Fig. 1(c) is the raised mid-section of the image, which is caused by retraction of the STM tip at low current. To quantify this height change, a line profile was extracted across the image closer to the slow scan direction, as shown in Fig. 1(e). The line location and direction is marked with an arrow to the immediate right of the image. The profile reveals that the tip pulls back approximately 0.45 nm to image at low current, an unusually large change in tip height. From other constant-current images taken (not shown), 0.05 nm of the height change must be attributed to the surface roughness of the copper. In addition, some tip retraction – about 0.10 nm for every factor of ten change in the current [31] – will naturally occur in STM as the current is reduced. However, these factors account for about half (0.25 nm) of the total height change (0.45 nm).

Only after variable-current simulated STM images were produced was the mechanism determined that explains the variable-current difference in the STM images. Six simulated images of graphene are displayed at the same scale in Fig. 2. Each image represents a surface of constant DOS, the value of which decreases from top to bottom. The highest DOS surface clearly shows the benzene ring structure, as seen on the left side of Fig. 2(a). In fact, the enormous hexagonally-shaped holes in the structure make up a majority of the entire surface area of the image, with individual carbon atoms clearly resolved. This is similar to STM data shown in the inset image of Fig. 1(a). At the medium DOS surface, the benzene ring loses almost all of its atomic-resolution features, as shown in Fig. 2(b). Note that the dark holes at the center of the benzene rings have shrunk considerably. At the lowest DOS surface, all of the atomic structural details disappear, as shown in Fig. 2(c). The image is characterized as a flat, featureless, uniform surface with small circular holes arranged throughout. From an STM perspective, these three top-view simulated images directly correlate to the experimental STM line profiles presented in Fig. 1(d). Note that as the tunneling current is decreased, the amplitude of the corrugations in the line profiles decreases correspondingly. Practically speaking, at very low currents, the ability to resolve the tiny circular depressions shown in the simulated images will be severely limited. Ultimately, resolution will be determined by the radius of curvature of the STM tip [32].

In addition to the top-view variable-current simulated STM images, side-view images were also created that reveal unusual extensions perpendicular to the surface. The highest constant DOS surface simulation is shown in the right side of Fig. 2(a). Here the total thickness of the electronic location of the graphene sample is about 0.2 nm, and the orbital structure above and below the carbon nucleus is clearly displayed. For the medium DOS surface, a dramatic change occurs as the isosurface shifts out into the vacuum an unusual amount, as shown in the right side of Fig. 2(b). The total thickness of the electronic location of the graphene layer is about 0.4 nm. Finally, for the lowest DOS simulation, the surface pushes out even further, as shown in the right side of Fig. 2(c). Here the thickness of the electronic location of the graphene is about 0.6 nm. The holes now appear as narrow columns penetrating from the surface to the nuclear plane below. From an STM perspective, the side-view images are comparable to the experimental STM line profile presented in Fig. 1(e). As the STM tunneling current is lowered by two orders of magnitude, the electronic location of the graphene surface moves out into the vacuum. The new graphene surface imaged by the tip would be about 0.2 nm higher than the original surface location [(0.6 nm − 0.2 nm)/2]. So, not only must the tip be farther away from the surface in order to achieve a lower current, but the surface itself is also higher. These two effects combine to cause the tip to retract approximately twice as far as would be expected.

To confirm that the above results are unique to graphene, additional extensive calculations were performed. A large number of variable-current simulated STM images were generated for both graphene and graphite by selecting incremental values for the constant DOS surfaces. For each of the simulated STM images (three of which are shown in Fig. 2), an average height (not the maximum height used earlier) was calculated. This data provided the integrated DOS (i.e., the charge density) as a function of average distance from the nuclear core for both graphene and graphite. These results are represented by the two curves in Fig. 3, and reveal that a massive difference exists. The highest charge density for graphene (upper curve) is 3 times greater than that of graphite (lower curve), despite the fact that they both peak at about one Bohr radius (0.05 nm). The peak location for graphene also represents a purely electronically derived corrugation, the theoretical value of which agrees quite well with the experimental corrugation shown in Fig. 1(b). These results also reveal how imaging charge densities with a magnitude accessible for graphite would cause the STM tip to be further from the surface for graphene. To illustrate, note the charge density circled on the lower curve near the Bohr radius, shown in Fig. 3. Now, imagine tunneling into graphite at this charge density when it suddenly becomes graphene. In response to this change, the STM tip would follow the dashed line away from the surface to the new point circled on the upper curve, just above the 0.1 nm mark.

As a final step in demonstrating the predicted differences, graphite was converted to graphene using the STM tip [33,34]. Data demonstrating this phenomenon is presented as an inset in Fig. 3, showing an STM image with two distinct regions. The left side of the STM image shows atomic-resolution graphite. About two-thirds of the way across the image, the surface suddenly changes to graphene. The graphite appears darker than the graphene section because it is lower in height. To see this more clearly, a line profile (location is marked with an arrow to the immediate left of the image) is shown directly below the STM image. While the STM tip was imaging the graphite part, its average height was 0.05 nm. When the surface switched to graphene, the tip was forced to pull back to an average height of about 0.20 nm for an overall height change of about 0.15 nm to maintain a constant current. The origin of this height change lies in Fig. 3, where the two circles are drawn. When the STM tip suddenly switches from imaging graphite to imaging graphene, the constant charge density surface extends into the vacuum more than 0.05 nm. This orbital extension shows that about one-third of the tip's total displacement is electronic, thus requiring the carbon atoms to move no more than 0.1 nm. Because graphite's carbon planes are weakly bonded, very little additional separation is necessary to "break" the bond. In testing this further, additional calculations were completed which show that the graphite surface atomic plane becomes a graphene surface after an additional inter-plane separation of about 0.1 nm, in excellent agreement with the STM data.

The primary discovery of this study is that variable-current topographical changes can be understood as an expansion of the topmost dangling bond orbitals of graphene (i.e., the $\pi$ orbitals) into the vacuum. For graphite, the charge density seeps into the bulk layers, resulting in the well-known triangular, rather than honeycomb, lattice [35,36]. For graphene, the absence of a bulk demands that the electron density extend into the vacuum in both directions, so that every atom is imaged, thereby contributing to the observed giant surface charge density. This large density of states naturally points to a large current carrying capacity [6,37,38] and allows STM at unusually high currents. Similarly, this provides an understanding of the basic origins of the superior current-carrying capacity and thermal conductivity of graphene.

In conclusion, this Rapid Communication presents STM images which resolve the atomic features associated with the giant surface charge density of graphene. The novel approach of producing variable-current experimental and first principles-generated images was used to uncover critical insights. These experimental and theoretical images are consistent in several key features which confirm reduced atomic resolution of the carbon atoms at lower tunneling currents. In addition, a clear understanding of the imaged features, in particular those taken using variable current, would not have been developed without the comparison to graphite. The theory and data corroborate in showing that carbon's filled-state $\pi$ orbitals become flat as they expand into the vacuum at low currents for graphene, but not graphite. This purely electronic expansion is due to the 300% larger surface charge density of graphene over graphite. Excellent agreement between first-principles theory and STM data in reproducing the electronic and geometric features emphasizes the need for this two-pronged approach in effectively evaluating atomic surface properties.

P.X. and P.T. are thankful for the financial support of the Office of Naval Research (ONR) under grant number N00014-10-1-0181 and the National Science Foundation (NSF) under grant number DMR-0855358. Y.Y. and L.B. thank the financial support of the Department of Energy, Office of Basic Energy Sciences, under contract ER-46612. They also acknowledge ONR Grants N00014-08-1-0915 and N00014-07-1-0825 (DURIP), and NSF grants DMR-0701558 and DMR-0080054 (C-SPIN). Some computations were made possible thanks to the MRI NSF grant 0722625 and to a Challenge grant from HPCMO of the U.S. Department of Defense.

∗thibado@uark.edu

**Figure Captions**

FIG. 1 (color online). (a) Filled-state STM image of graphene on copper foil acquired with a tip bias of +0.100 V and a tunneling current of 1.000 nA measuring 6 nm × 6 nm. Inset: STM image cut from the center of 1(a) magnified 2.5 times, and displayed with a compressed color scale to reveal the individual atomic orbitals, (b) height cross section line profile extracted from the STM image shown in (a) across the horizontal row marked with the arrow just to the right of the image, (c) filled-state STM image of graphene on copper foil acquired with a tip bias of +0.100 V and a tunneling current ranging from 0.010 nA (L) to 1.000 nA (H) with steps in between (M) and measuring 6 nm × 6 nm, (d) offset height cross sections line profiles extracted from three different regions of the STM image shown in (c) across vertical rows, and (e) height cross section line profile extracted from the STM image shown in (c) and marked with the arrow just to the upper right of the image.

FIG. 2. Simulated STM images of graphene using a filled-state bias of 0.01 eV below the Dirac point as shown in a top view (left) and side view (right). Each represents a surface of constant DOS, the isovalue is noted.

FIG. 3 (color online). Plot of the spatial charge density as a function of distance from the nuclear plane for graphene (upper curve) and graphite (lower curve) calculated from simulated constant DOS images. (Inset) Filled-state STM image of HOPG acquired with a tip bias of +0.100 V and a tunneling current of 0.200 nA measuring 6 nm × 6 nm. Below the image is a height cross section that was extracted from the image along the line marked with the arrow just to the left of the image.

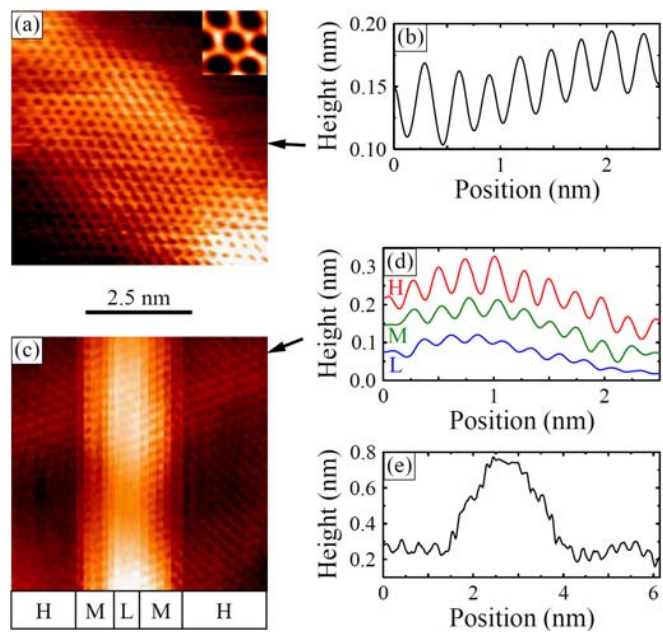

FIG. 1    by P. Xu *et al.*

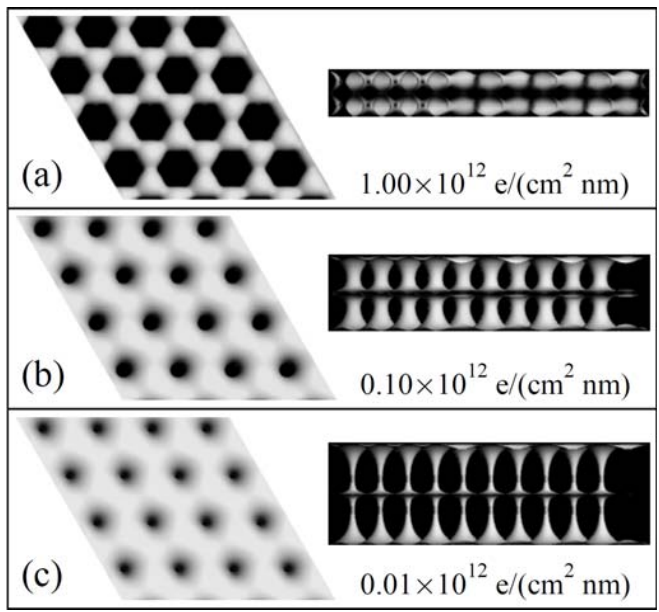

FIG. 2 by P. Xu *et al.*

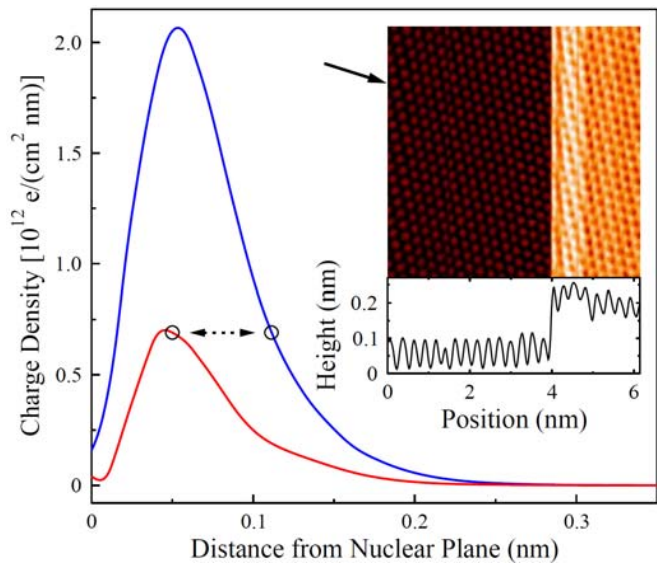

FIG. 3    by P. Xu *et al.*